\documentclass[%
 aip,
 amsmath,amssymb, 
 reprint,longbibliography
]{revtex4-1}
\usepackage{graphicx}
\draft 

\begin{document}

\title{Cool Cooling Collar for Bake-Out of Temperature-Sensitive Devices}

\author{\firstname{Jonas D.} Fortmann}
\email[Corresponding author: ]{jonas.fortmann@uni-due.de}
\affiliation{Faculty of Physics, University of Duisburg-Essen, 47057 Duisburg, Germany}

\author{Christian Brand}
\affiliation{Faculty of Physics, University of Duisburg-Essen, 47057 Duisburg, Germany}

\author{Michael \surname{Horn-von Hoegen}}
\affiliation{Faculty of Physics, University of Duisburg-Essen, 47057 Duisburg, Germany}
\affiliation{Center for Nanointegration (CENIDE), University of Duisburg-Essen, 47057 Duisburg, Germany}

\date{\today}

\begin{abstract}
A combination of a pumpable gate valve and a self-built cooling collar permits bake-out of an ultra-high vacuum chamber  without having to dismount sensitive equipment. 
A small pump port on the closed gate valve maintains ultra-high vacuum conditions for a TVIPS TemCam-XF416 imaging electron detector in the case of venting the main chamber. 
The water-cooled collar mounted to the detector housing prevents heating of the detector upon bake-out of the ultra-high vacuum chamber.
\end{abstract}

\maketitle

\section{Introduction}

With the commercial availability of modern imaging electron detectors with high pixel resolution for application in transmission electron microscopes, there is growing interest in using these cameras for experiments under ultra-high vacuum (UHV) conditions.
These fiber-optically coupled CMOS-based detectors are fully integrated with a thin polycrystalline phosphor scintillator as the electron-sensitive component. 
Under all circumstances, such detectors must not become hotter than 40°C, otherwise serious damage may occur.
However, establishing UHV conditions requires prolonged bake-out of the entire chamber to temperatures well above 100°C which is incompatible with the maximum temperature requirements of the camera.  
A possible solution may be to dismount the detector before bake-out but involves considerable drawbacks.
The detector can be physically damaged during assembly and reassembly.
It is exposed to ambient conditions and moisture that enter the UHV chamber after reconnecting the detector. 
Upon dismounting and re-attaching the detector, the warming-up and cooling-down cycles introduce thermal stress to the fiber-CMOS unit that should be avoided.

Here, we report on the upgrade of our ultra-fast electron diffraction experiment \cite{janzenUltrafastElectronDiffraction2006,
hanisch-blicharskiUltrafastElectronDiffraction2013,
frigge_optically_2017,
hafke_pulsed_2019,
hanisch-blicharski_violation_2021,
horn-vonhoegenInvitedReview2023}
from a multi channel plate detector to a fiber-optically coupled CMOS-based detector (TVIPS TemCam-XF416) \cite{tietzAdvantagesFiberOpticallyCMOS2022}.
With the help of a self-built water-cooled collar we cool the housing of the detector and thus permit bake-out of the entire chamber without dismounting and warming up of the detector.  

\section{Setup}

\begin{figure}[htbp]
\centering
\includegraphics[width=1.000\columnwidth]{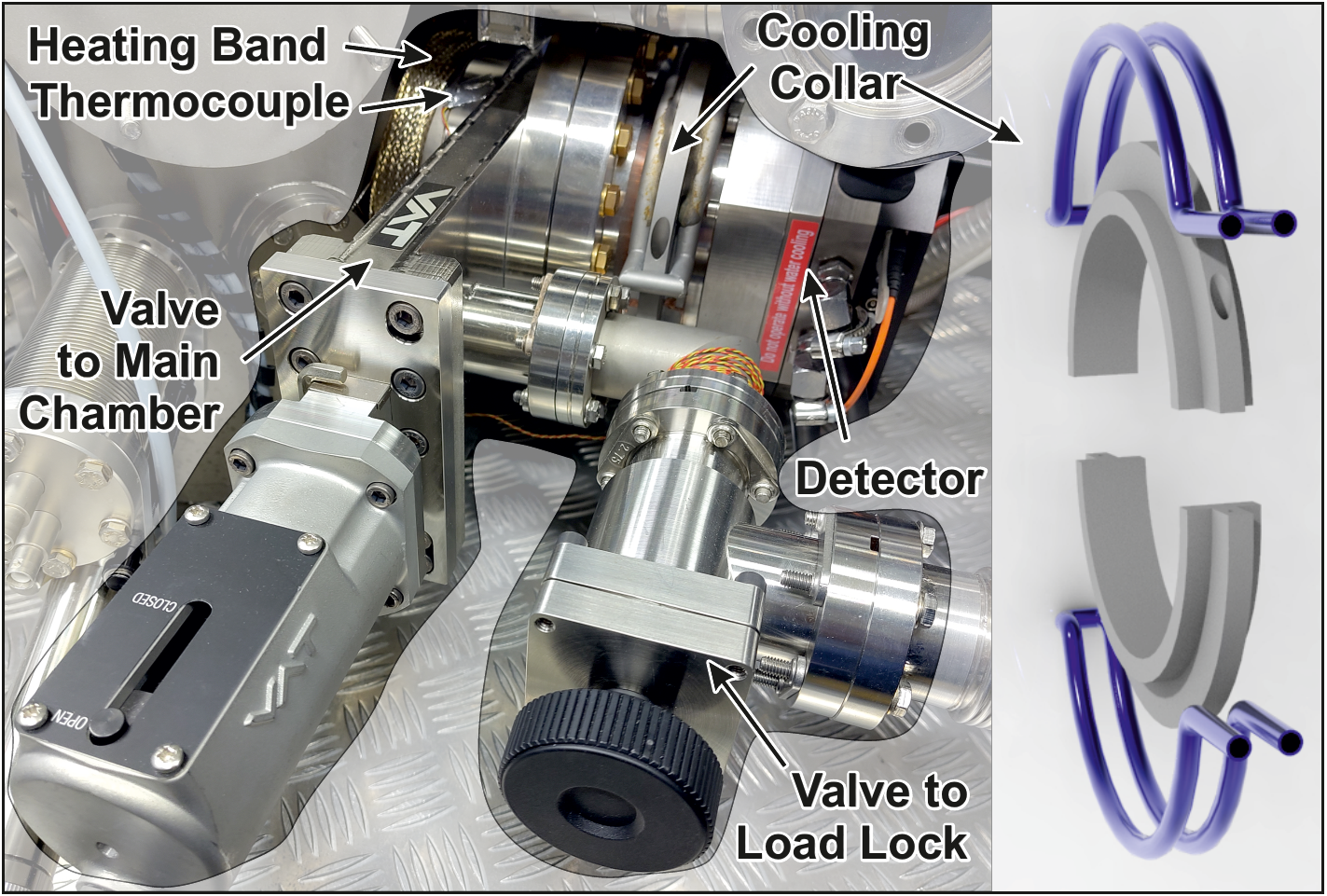}
\caption{
\textbf{Our Setup} depicting the cooling collar (exploded view in the right panel), attached directly on the TVIPS TemCam-XF416 detector, which is mounted via a VAT gate valve CF-F 100 CF-F 160 at the main chamber. On the detector side an additional CF-F 40 pump flange is connected to our load lock system through an angle valve. On the side of the main chamber a heating tape is located and a type K thermocouple is installed in order to monitor the temperature.
}
\label{fig:Setup}
\end{figure}
In our setup shown in Fig.\ref{fig:Setup}, the detector is mounted to the main chamber (MC) via a gate valve (VAT CF-F 100 CF-F 160).
The gate valve exhibits an additional CF-F 40 flange on the detector side with an angle valve (VAT 28.4 CF-F 40) attached, which is connected to a turbo molecular pump (Pfeiffer HiPace 80) at the load lock chamber (LL) with a base pressure better than 3e-8\,mbar.
This combination allows the detector to be pumped through the MC or connected to the LL without mutual interference.
During imaging operation of the detector the angle valve to the LL is closed.
The MC of the tr-RHEED experiment is pumped by a turbo molecular pump (Pfeiffer TMU 521), an ion getter pump (Varian) and two getter pumps (SAES CapaciTorr\textregistered-D 400-2).
The pressure is measured by means of an ion gauge (MKS Granville Phillips GP307).

In case of venting the MC the gate valve is closed, the detector is kept cold via its internal Peltier cooler, and is pumped by the LL. Exposure to ambient conditions is thus avoided.
While the gate valve is heated during bake-out to ensure that UHV conditions are reached, the detector is located outside the bake-out oven and is not heated by radiation or convection.
During bake-out, the gate valve is closed and thus protects the szintillator surface from any direct heat radiation. 

To prevent heating of the detector through thermal conductivity via the hot gate valve we mounted a water cooled collar on the detector housing acting as heat sink.  
This collar is made of two parts of solid stainless steal half-rings, as shown in Fig.\ref{fig:Setup}. 
Each half collar is 30\,mm wide and 10\,mm thick. They are cooled by a capillary (stainless steal tube of 10\,mm outer and 9\,mm inner diameter), hard soldered for maximizing the thermal conductivity. 
Both capillaries are connected with PVC fibre-reinforced hose pipes in one single loop to the cooling water (15°C, 1.5\,l/min).

The inside of each half collar is covered with heat conducting copper tape to maximize the thermal contact across the 139.5\,cm$^2$ contact area to the detector housing.
Both collar half rings are tightly connected to each other through two M4 screws (tightened at 4\,Nm torque) and thus pressed onto the detector housing.

\section{Results}

\begin{figure}[htbp]
\centering
\includegraphics[width=1.000\columnwidth]{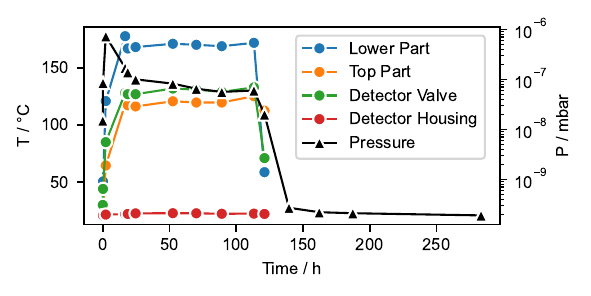} 
\caption{
\textbf{Temperature and pressure evolution during bake-out}. The temperature of the gate valve separating MC and detector was measured on the flange towards the MC. 
}
\label{fig:Baek_Data}
\end{figure}

Using type K thermocouples located at various positions of the MC the temperatures during bake-out were monitored and are shown in Fig. \ref{fig:Baek_Data}. 
The temperature of the detector housing (red data points and sold line) rises just one Kelvin from 21°C to 22°C while the temperature of the gate valve connected to the detector increases up to 130°C. The low temperature rise by 1\,K proves the efficiency of our cooling collar.
The heating of the lower MC together with the IGP and turbo pump was turned off after 5 days, while the upper part of the MC followed 12 hours later.
Subsequently, the pressures dropped from 6.0 e-8\,mbar to below 2e-10 mbar (black triangles and line in Fig.\ref{fig:Baek_Data}.
The detector was reconnected to the MC by opening of the gate valve and closing the angle valve to the LL one day after bake-out.

\section{Conclusions}
With the help of a water-cooled cooling collar the bake-out of an UHV chamber with sensitive equipment like an imaging electron detector is possible without dismounting. 
In our case, the initial first pump down of the detector required three weeks until UHV conditions were reached.
After that, the detector never needs to be vented again.
After venting the MC for maintenance and subsequent bakeout, UHV conditions with $p < 2$e-10\,mbar were routinely achieved.

\section*{Acknowledgments}
Fruitful discussions with TVIPS are gratefully acknowledged. Funded by the Deutsche Forschungsgemeinschaft (DFG, German Research Foundation) through projects B04 and C03 of Collaborative Research Center SFB1242 ``Nonequilibrium dynamics of condensed matter in the time domain'' (Project-ID 278162697).

\section*{Author Declaration}

\subsection{Conflict of Interest}
The authors have no conflicts to disclose.

\subsection{Author Contributions}
\textbf{Jonas D. Fortmann}:
Conceptualization (equal);
Data curation;
Visualization;
Writing – original draft;
Writing – review \& editing (equal).
\textbf{Christian Brand}:
Conceptualization (equal);
Construction;
Supervision (equal);
Writing – review \& editing (supporting).
\textbf{Michael Horn-von Hoegen}:
Conceptualization (supporting);
Supervision (equal);
Writing – original draft (supporting);
Writing – review \& editing (equal).

\section*{Data Availability}

The data that supports the findings of this study are available from the corresponding author upon reasonable request.

\section*{References}
\bibliography{Cooler}

\begin{thebibliography}{7}%
\makeatletter
\providecommand \@ifxundefined [1]{%
 \@ifx{#1\undefined}
}%
\providecommand \@ifnum [1]{%
 \ifnum #1\expandafter \@firstoftwo
 \else \expandafter \@secondoftwo
 \fi
}%
\providecommand \@ifx [1]{%
 \ifx #1\expandafter \@firstoftwo
 \else \expandafter \@secondoftwo
 \fi
}%
\providecommand \natexlab [1]{#1}%
\providecommand \enquote  [1]{``#1''}%
\providecommand \bibnamefont  [1]{#1}%
\providecommand \bibfnamefont [1]{#1}%
\providecommand \citenamefont [1]{#1}%
\providecommand \href@noop [0]{\@secondoftwo}%
\providecommand \href [0]{\begingroup \@sanitize@url \@href}%
\providecommand \@href[1]{\@@startlink{#1}\@@href}%
\providecommand \@@href[1]{\endgroup#1\@@endlink}%
\providecommand \@sanitize@url [0]{\catcode `\\12\catcode `\$12\catcode `\&12\catcode `\#12\catcode `\^12\catcode `\_12\catcode `\%12\relax}%
\providecommand \@@startlink[1]{}%
\providecommand \@@endlink[0]{}%
\providecommand \url  [0]{\begingroup\@sanitize@url \@url }%
\providecommand \@url [1]{\endgroup\@href {#1}{\urlprefix }}%
\providecommand \urlprefix  [0]{URL }%
\providecommand \Eprint [0]{\href }%
\providecommand \doibase [0]{http://dx.doi.org/}%
\providecommand \selectlanguage [0]{\@gobble}%
\providecommand \bibinfo  [0]{\@secondoftwo}%
\providecommand \bibfield  [0]{\@secondoftwo}%
\providecommand \translation [1]{[#1]}%
\providecommand \BibitemOpen [0]{}%
\providecommand \bibitemStop [0]{}%
\providecommand \bibitemNoStop [0]{.\EOS\space}%
\providecommand \EOS [0]{\spacefactor3000\relax}%
\providecommand \BibitemShut  [1]{\csname bibitem#1\endcsname}%
\let\auto@bib@innerbib\@empty
\bibitem [{\citenamefont {Janzen}\ \emph {et~al.}(2006)\citenamefont {Janzen}, \citenamefont {Krenzer}, \citenamefont {Zhou}, \citenamefont {{von der Linde}},\ and\ \citenamefont {{Horn-von Hoegen}}}]{janzenUltrafastElectronDiffraction2006}%
  \BibitemOpen
  \bibfield  {author} {\bibinfo {author} {\bibfnamefont {A.}~\bibnamefont {Janzen}}, \bibinfo {author} {\bibfnamefont {B.}~\bibnamefont {Krenzer}}, \bibinfo {author} {\bibfnamefont {P.}~\bibnamefont {Zhou}}, \bibinfo {author} {\bibfnamefont {D.}~\bibnamefont {{von der Linde}}}, \ and\ \bibinfo {author} {\bibfnamefont {M.}~\bibnamefont {{Horn-von Hoegen}}},\ }\href {\doibase 10.1016/j.susc.2006.02.070} {\bibfield  {journal} {\bibinfo  {journal} {Surface Science}\ }\textbf {\bibinfo {volume} {600}},\ \bibinfo {pages} {4094} (\bibinfo {year} {2006})}\BibitemShut {NoStop}%
\bibitem [{\citenamefont {{Hanisch-Blicharski}}\ \emph {et~al.}(2013)\citenamefont {{Hanisch-Blicharski}}, \citenamefont {Janzen}, \citenamefont {Krenzer}, \citenamefont {Wall}, \citenamefont {Klasing}, \citenamefont {Kalus}, \citenamefont {Frigge}, \citenamefont {Kammler},\ and\ \citenamefont {{Horn-von Hoegen}}}]{hanisch-blicharskiUltrafastElectronDiffraction2013}%
  \BibitemOpen
  \bibfield  {author} {\bibinfo {author} {\bibfnamefont {A.}~\bibnamefont {{Hanisch-Blicharski}}}, \bibinfo {author} {\bibfnamefont {A.}~\bibnamefont {Janzen}}, \bibinfo {author} {\bibfnamefont {B.}~\bibnamefont {Krenzer}}, \bibinfo {author} {\bibfnamefont {S.}~\bibnamefont {Wall}}, \bibinfo {author} {\bibfnamefont {F.}~\bibnamefont {Klasing}}, \bibinfo {author} {\bibfnamefont {A.}~\bibnamefont {Kalus}}, \bibinfo {author} {\bibfnamefont {T.}~\bibnamefont {Frigge}}, \bibinfo {author} {\bibfnamefont {M.}~\bibnamefont {Kammler}}, \ and\ \bibinfo {author} {\bibfnamefont {M.}~\bibnamefont {{Horn-von Hoegen}}},\ }\href {\doibase 10.1016/j.ultramic.2012.07.017} {\bibfield  {journal} {\bibinfo  {journal} {Ultramicroscopy}\ }\textbf {\bibinfo {volume} {127}},\ \bibinfo {pages} {2} (\bibinfo {year} {2013})}\BibitemShut {NoStop}%
\bibitem [{\citenamefont {Frigge}\ \emph {et~al.}(2017)\citenamefont {Frigge}, \citenamefont {Hafke}, \citenamefont {Witte}, \citenamefont {Krenzer}, \citenamefont {Streub{\"u}hr}, \citenamefont {Samad~Syed}, \citenamefont {Mik{\v s}i{\'c}~Trontl}, \citenamefont {Avigo}, \citenamefont {Zhou}, \citenamefont {Ligges}, \citenamefont {Von Der~Linde}, \citenamefont {Bovensiepen}, \citenamefont {{Horn-von Hoegen}}, \citenamefont {Wippermann}, \citenamefont {L{\"u}cke}, \citenamefont {Sanna}, \citenamefont {Gerstmann},\ and\ \citenamefont {Schmidt}}]{frigge_optically_2017}%
  \BibitemOpen
  \bibfield  {author} {\bibinfo {author} {\bibfnamefont {T.}~\bibnamefont {Frigge}}, \bibinfo {author} {\bibfnamefont {B.}~\bibnamefont {Hafke}}, \bibinfo {author} {\bibfnamefont {T.}~\bibnamefont {Witte}}, \bibinfo {author} {\bibfnamefont {B.}~\bibnamefont {Krenzer}}, \bibinfo {author} {\bibfnamefont {C.}~\bibnamefont {Streub{\"u}hr}}, \bibinfo {author} {\bibfnamefont {A.}~\bibnamefont {Samad~Syed}}, \bibinfo {author} {\bibfnamefont {V.}~\bibnamefont {Mik{\v s}i{\'c}~Trontl}}, \bibinfo {author} {\bibfnamefont {I.}~\bibnamefont {Avigo}}, \bibinfo {author} {\bibfnamefont {P.}~\bibnamefont {Zhou}}, \bibinfo {author} {\bibfnamefont {M.}~\bibnamefont {Ligges}}, \bibinfo {author} {\bibfnamefont {D.}~\bibnamefont {Von Der~Linde}}, \bibinfo {author} {\bibfnamefont {U.}~\bibnamefont {Bovensiepen}}, \bibinfo {author} {\bibfnamefont {M.}~\bibnamefont {{Horn-von Hoegen}}}, \bibinfo {author} {\bibfnamefont {S.}~\bibnamefont {Wippermann}}, \bibinfo {author} {\bibfnamefont {A.}~\bibnamefont {L{\"u}cke}}, \bibinfo {author}
  {\bibfnamefont {S.}~\bibnamefont {Sanna}}, \bibinfo {author} {\bibfnamefont {U.}~\bibnamefont {Gerstmann}}, \ and\ \bibinfo {author} {\bibfnamefont {W.~G.}\ \bibnamefont {Schmidt}},\ }\href {\doibase 10.1038/nature21432} {\bibfield  {journal} {\bibinfo  {journal} {Nature}\ }\textbf {\bibinfo {volume} {544}},\ \bibinfo {pages} {207} (\bibinfo {year} {2017})}\BibitemShut {NoStop}%
\bibitem [{\citenamefont {Hafke}\ \emph {et~al.}(2019)\citenamefont {Hafke}, \citenamefont {Witte}, \citenamefont {Brand}, \citenamefont {Duden},\ and\ \citenamefont {{Horn-von Hoegen}}}]{hafke_pulsed_2019}%
  \BibitemOpen
  \bibfield  {author} {\bibinfo {author} {\bibfnamefont {B.}~\bibnamefont {Hafke}}, \bibinfo {author} {\bibfnamefont {T.}~\bibnamefont {Witte}}, \bibinfo {author} {\bibfnamefont {C.}~\bibnamefont {Brand}}, \bibinfo {author} {\bibfnamefont {{\relax Th}.}~\bibnamefont {Duden}}, \ and\ \bibinfo {author} {\bibfnamefont {M.}~\bibnamefont {{Horn-von Hoegen}}},\ }\href {\doibase 10.1063/1.5086124} {\bibfield  {journal} {\bibinfo  {journal} {Review of Scientific Instruments}\ }\textbf {\bibinfo {volume} {90}},\ \bibinfo {pages} {045119} (\bibinfo {year} {2019})}\BibitemShut {NoStop}%
\bibitem [{\citenamefont {{Hanisch-Blicharski}}\ \emph {et~al.}(2021)\citenamefont {{Hanisch-Blicharski}}, \citenamefont {Tinnemann}, \citenamefont {Wall}, \citenamefont {Thiemann}, \citenamefont {Groven}, \citenamefont {Fortmann}, \citenamefont {Tajik}, \citenamefont {Brand}, \citenamefont {Frost}, \citenamefont {Von~Hoegen},\ and\ \citenamefont {{Horn-von Hoegen}}}]{hanisch-blicharski_violation_2021}%
  \BibitemOpen
  \bibfield  {author} {\bibinfo {author} {\bibfnamefont {A.}~\bibnamefont {{Hanisch-Blicharski}}}, \bibinfo {author} {\bibfnamefont {V.}~\bibnamefont {Tinnemann}}, \bibinfo {author} {\bibfnamefont {S.}~\bibnamefont {Wall}}, \bibinfo {author} {\bibfnamefont {F.}~\bibnamefont {Thiemann}}, \bibinfo {author} {\bibfnamefont {T.}~\bibnamefont {Groven}}, \bibinfo {author} {\bibfnamefont {J.}~\bibnamefont {Fortmann}}, \bibinfo {author} {\bibfnamefont {M.}~\bibnamefont {Tajik}}, \bibinfo {author} {\bibfnamefont {C.}~\bibnamefont {Brand}}, \bibinfo {author} {\bibfnamefont {B.-O.}\ \bibnamefont {Frost}}, \bibinfo {author} {\bibfnamefont {A.}~\bibnamefont {Von~Hoegen}}, \ and\ \bibinfo {author} {\bibfnamefont {M.}~\bibnamefont {{Horn-von Hoegen}}},\ }\href {\doibase 10.1021/acs.nanolett.1c01665} {\bibfield  {journal} {\bibinfo  {journal} {Nano Letters}\ }\textbf {\bibinfo {volume} {21}},\ \bibinfo {pages} {7145} (\bibinfo {year} {2021})}\BibitemShut {NoStop}%
\bibitem [{\citenamefont {{Horn-von Hoegen}}(2023)}]{horn-vonhoegenInvitedReview2023}%
  \BibitemOpen
  \bibfield  {author} {\bibinfo {author} {\bibfnamefont {M.}~\bibnamefont {{Horn-von Hoegen}}},\ }\href@noop {} {\bibfield  {journal} {\bibinfo  {journal} {Structrual Dynamics}\ }\textbf {\bibinfo {volume} {submitted}} (\bibinfo {year} {2023})}\BibitemShut {NoStop}%
\bibitem [{\citenamefont {Tietz}\ \emph {et~al.}(2022)\citenamefont {Tietz}, \citenamefont {Oster}, \citenamefont {Wisnet},\ and\ \citenamefont {Tietz}}]{tietzAdvantagesFiberOpticallyCMOS2022}%
  \BibitemOpen
  \bibfield  {author} {\bibinfo {author} {\bibfnamefont {H.}~\bibnamefont {Tietz}}, \bibinfo {author} {\bibfnamefont {M.}~\bibnamefont {Oster}}, \bibinfo {author} {\bibfnamefont {A.}~\bibnamefont {Wisnet}}, \ and\ \bibinfo {author} {\bibfnamefont {D.}~\bibnamefont {Tietz}},\ }\href@noop {} {\enquote {\bibinfo {title} {Advantages of {{Fiber-Optically CMOS Cameras}} for {{LEEM}}/{{PEEM Applications}}},}\ } (\bibinfo {year} {2022}),\ \bibinfo {note} {12th {{International Conference}} on {{LEEM}} – {{PEEM}}}\BibitemShut {NoStop}%
\end{thebibliography}%

Copyright (2023) Jonas D. Fortmann, Christian Brand \& Michael Horn-von Hoegen. This article is distributed under a Creative Commons Attribution (CC BY-NC-ND) License

\end{document}